\documentclass[prb,twocolumn,showkeys,amssymb,floatfix]{revtex4}

\usepackage{cancel}
\usepackage{color}
\usepackage{graphicx}

\usepackage{mathrsfs} 

\usepackage{empheq}
\usepackage{rbraket} 

\usepackage{amsmath}
\usepackage{amsthm}

\DeclareMathAlphabet{\mathtensor}{OT1}{cmss}{bx}{n}

\theoremstyle{plain}

\theoremstyle{definition}

\begin{document}


\title{Mori-Zwanzig Equations With Time-Dependent Liouvillian}

\author{Louis-S. Bouchard}
 \affiliation{Materials Sciences Division, Lawrence Berkeley National
 Laboratory and Department of Chemistry, University of California,
 Berkeley, CA 94720}
 \homepage{http://waugh.cchem.berkeley.edu}

\date{\today}

\begin{abstract}
We extend the Mori-Zwanzig equations to the case where the Liouvillian
is time-dependent by way of a Dyson identity due to Holian and
Evans. This extension makes it possible to treat, for example, more general
cases of continuous-wave (cw) or pulsed irradiation in NMR and
coherent optics. The formalism immediately allows for reference frame
changes and approximations based on coherent averaging.
\end{abstract}

\pacs{76.60.Jx}

\keywords{nuclear induction, non-equilibrium statistical mechanics,
coherent optics, coherent averaging, spin relaxation, atomic
relaxation, mori-zwanzig theory, mori theory}

\maketitle

\newfont{\Bb}{msbm10}

\newcommand{\dotprod}{{\scriptscriptstyle \stackrel{\bullet}{{}}}}

\section{Introduction}

Liouville space theories such as the Mori-Zwanzig theory may be used
to describe the evolution of a set of quantum mechanical
observables~\cite{bib:fick_sauermann} under conditions of
non-equilibria. To the author's knowledge, the current formalism 
published in the nuclear magnetic resonance (NMR)
literature~\cite{bib:mehring_1976book,bib:fick_sauermann,bib:abragam_goldman}
does not include the possibility for time-dependent Hamiltonian
dynamics~\footnote{A mere replacement of the time-independent
Liouvillian by a time-dependent Liouvillian is not possible unless 
further justification is provided.}. The usual quantum mechanical
master equation obtained from a perturbation approach~\cite{bib:abragam} is
the tool of choice for describing time-dependent
dynamics~\cite{bib:rhim,bib:levittdibari}. These methods normally rely
on the assumption of a short correlation time.

The Mori equations provide a convenient framework for describing
time-evolution in more general and not necessarily perturbative 
cases, either by analytical or numerical solution. Modern experiments
increasingly consist of rapid stochastic or periodic
irradiations~\cite{bib:rhim,bib:levittdibari,bib:zewail}
for which the timescale of the applied pulses may approach that of
molecular motions~\cite{bib:zewail}. 
It is therefore sensible to extend the current formalism so these
modern experiments can be described adequately.

We begin by summarizing the usual derivation of the Mori equations for
time-independent Liouvillian dynamics. We then derive the equations in
the case case of time-dependent Liouvillian.


\section{ Time-Independent Hamiltonians }

\subsection{Dyson Formula}

The usual derivation of the Mori equations starts from the Dyson
formula~\cite{bib:fick_sauermann,bib:zwanzig_book,bib:kivelson}, the
derivation of which is repeated here for convenience. Differentiation
of the function

\begin{equation}
\mathtensor{A}(s) = \mathrm{e}^{-s \mathtensor{L}} \mathrm{e}^{
  s(1-\pi)  \mathtensor{L} }
\label{eq:fofs}
\end{equation}

\noindent with respect to $s$ gives:

\begin{equation}
\frac{ d \mathtensor{A}(s) }{ \mathrm{d}s } = - \mathrm{e}^{ -s
  \mathtensor{L} } \pi \mathtensor{L} \mathrm{e}^{s (1-\pi)  \mathtensor{L}},
\end{equation}

\noindent where $\mathtensor{L}$ is independent of $s$.
Integration from $0$ to $t$ leads to:

\begin{equation}
\mathtensor{A}(t) = \mathtensor{A}(0) - \int_0^t \mathrm{e}^{ -s
\mathtensor{L} } \pi \mathtensor{L} \mathrm{e}^{s (1-\pi)
  \mathtensor{L}} \mathrm{d}s.
\end{equation}

\noindent Substitution of Eq.(\ref{eq:fofs}) and
$\mathtensor{A}(0)=\mathtensor{1}$, and multiplication on the left by
$\mathrm{e}^{t \mathtensor{L}}$ yields the desired result:

\begin{equation}
\mathrm{e}^{t(1-\pi) \mathtensor{L}} = \mathrm{e}^{t \mathtensor{L} } - \int_0^t
\mathrm{e}^{(t-s) \mathtensor{L}} \pi \mathtensor{L} \mathrm{e}^{s (1-\pi)
\mathtensor{L} } \mathrm{d}s.
\label{eq:dyson1}
\end{equation}

\subsection{Heisenberg Equation of Motion}

The Heisenberg equation of motion for an
observable $A(t)$:

\begin{equation}
\frac{d}{dt} A(t) = \mathtensor{L} A(t)
\label{eq:eq1}
\end{equation}

\noindent has the following solution

\begin{equation}
A(t) = \mathrm{e}^{t \mathtensor{L}} A(0).
\end{equation}

\noindent where $\mathtensor{L}$ is the time-independent Liouvillian,
defined by the commutator $\mathtensor{L}A=i[H,A]$.

\subsection{Matrix Operator Notation}

Consider set of $n$ quantum operators $\{ I_1, I_2, \dots I_n \}$ and
write them as a column vector 

\begin{equation}
 \overrightarrow{ \mathscr{G} } = \left[
\begin{matrix}
I_1 \\
I_2 \\
\vdots \\
I_n
\end{matrix} \right]. 
\end{equation}

\noindent To this column vector corresponds a row vector:

\begin{equation}
\overleftarrow{ \mathscr{G} } = \left[
\begin{matrix}
I_1 & I_2 & \dots & I_n
\end{matrix} \right].
\end{equation}

Next, an inner product is chosen which is most appropriate to the
problem under study. The notation $\rbraket{ A | B} $ denotes the
inner product of $A$ and $B$. There are several popular inner products
which may be used, depending on the application:

\begin{enumerate}
\item The trace projection~\cite{bib:ernstbook,bib:jeenersuperop},
also called the Hilbert-Schmidt inner product:

\begin{equation}
\rbraket{ A | B}_{t.p.} = \mbox{Tr} \left[ A^\dagger B \right].
\label{eq:ip1}
\end{equation}

\item The thermal average
~\cite{bib:fick_sauermann,bib:kivelson}, which depends on the
temperature $\beta=1/kT$ and Hamiltonian $H$ of the system:

\begin{equation}
\rbraket{ A | B}_{t.a.} = \mbox{Tr} \left[ A^\dagger B \frac{ \mathrm{e}^{-
 \beta H } }{Z} \right]; Z=\mbox{Tr} \left[ \mathrm{e}^{-\beta H} \right],
\label{eq:ip2}
\end{equation}

\noindent where $\exp(-\beta H)/Z$ is the canonical statistical
operator.

\item The Mori product, which also depends on $\beta$ and $H$:

\begin{equation}
\rbraket{ A | B }_{m.p.} =  \frac{1}{\beta}  \int_0^{\beta}
  \mbox{Tr} \left[ \frac{ e^{-\beta H} }{Z} A^\dagger 
  e^{-u \beta H} B e^{ u \beta H} \right] \mathrm{d}u.
\label{eq:ip3}
\end{equation}

\end{enumerate}

Next, we define the following projection superoperator,

\begin{equation}
\pi \rket{ \overrightarrow{ \mathscr{F} } } = \rbraket{ \overrightarrow{\mathscr{F}} |
    \overleftarrow{\mathscr{G} }} \cdot \rbraket{ \overrightarrow{\mathscr{G}} |
    \overleftarrow{\mathscr{G} }}^{-1} \rket{ \overrightarrow{
    \mathscr{G} } }.
\end{equation}

\noindent In component form, this is:

\begin{equation}
\pi \rket{ \mathscr{F}_i } = \sum_{j,k} \rbraket{ \mathscr{F}_i | \mathscr{G}_j }
 \left[ \rbraket{ \overrightarrow{ \mathscr{G} } |
 \overleftarrow{\mathscr{G} }}^{-1}
 \right]_{jk} \rket{ \mathscr{G}_k }.
\end{equation}

The symbol $\rbraket{ \overrightarrow{ \mathscr{G} } |
\overleftarrow{\mathscr{G} }}$ denotes the $n$-by-$n$ matrix formed by outer
product of the column vector $\rbra{ \overrightarrow{ \mathscr{G} }
}$ and the row vector $\rket{ \overleftarrow{\mathscr{G} }}$:

\begin{widetext}
\begin{equation}
 \rbraket{ \overrightarrow{ \mathscr{G} } |
 \overleftarrow{\mathscr{G} }}
= ( \left[
\begin{matrix}
I_1 \\
I_2 \\
\vdots \\
I_n
\end{matrix} \right] | \left[
\begin{matrix}
I_1 & I_2 & \dots & I_n
\end{matrix} \right] ) = 
\left[ \begin{matrix}
\rbraket{I_1 | I_1} & \rbraket{I_1 | I_2} & \dots & \rbraket{I_1 | I_n} \\
\rbraket{I_2 | I_1} & \rbraket{I_2 | I_2} & \dots & \rbraket{I_2 | I_n} \\
\vdots \\
\rbraket{I_n | I_1} & \rbraket{I_n | I_2} & \dots & \rbraket{I_n | I_n} 
\end{matrix} \right]
\label{eq:outerp}
\end{equation}
\end{widetext}

\noindent whose entries are the inner products $\rbraket{I_i | I_j}$, 
whereas 

$$\rbraket{ \overrightarrow{ \mathscr{G} } |
 \overleftarrow{\mathscr{G} }}^{-1}$$

\noindent is the inverse of that matrix.
Thus, a superoperator commutes with $\rbraket{ \overrightarrow{
\mathscr{G} } |  \overleftarrow{\mathscr{G} }}^{-1}$ and $\rbraket{
\overrightarrow{ \mathscr{G} } | \overleftarrow{\mathscr{G} }}$
because the entries of those matrices are $c$-numbers.

When the bracket is acting on a matrix $\overrightarrow{
\mathscr{F} } \overleftarrow{ \mathscr{G} }$, the
inner product is taken over all entries of the matrix individually, as
was done in Eq.(\ref{eq:outerp}).

We check that this is indeed a valid projection operator:

\begin{align*}
\pi^2 \rket{ \overrightarrow{ \mathscr{F} } } =& \rbraket{
  ( \overrightarrow{\mathscr{F}} |
    \overleftarrow{\mathscr{G} } ) \cdot ( \overrightarrow{\mathscr{G}} |
    \overleftarrow{\mathscr{G} } ) ^{-1} \overrightarrow{ \mathscr{G} } |
    \overleftarrow{\mathscr{G} }} 
\cdot \rbraket{ \overrightarrow{\mathscr{G}} |
    \overleftarrow{\mathscr{G} }}^{-1} \rket{ \overrightarrow{
    \mathscr{G} } } \\
=& \rbraket{ \overrightarrow{\mathscr{F}} |
    \overleftarrow{\mathscr{G} } } \cdot \rbraket{ \overrightarrow{\mathscr{G}} |
    \overleftarrow{\mathscr{G} } } ^{-1}  \rbraket{  \overrightarrow{
  \mathscr{G} } |
    \overleftarrow{\mathscr{G} }} 
\cdot \rbraket{ \overrightarrow{\mathscr{G}} |
    \overleftarrow{\mathscr{G} }}^{-1} \rket{ \overrightarrow{
    \mathscr{G} } } \\
=& \rbraket{ \overrightarrow{\mathscr{F}} |
    \overleftarrow{\mathscr{G} } } 
\cdot \rbraket{ \overrightarrow{\mathscr{G}} |
    \overleftarrow{\mathscr{G} }}^{-1} \rket{ \overrightarrow{
    \mathscr{G} } } = \pi \rket{ \overrightarrow{ \mathscr{F} } }.
\end{align*}

The form of the projector is particularly simple if the set of
operators is orthogonal. For example, with $\{I_+ , I_-, I_z\}$ the
matrix $\rbraket{ \overrightarrow{ \mathscr{G} } |
 \overleftarrow{\mathscr{G} }}$ is diagonal:

\begin{equation}
 ( \left[
\begin{matrix}
I_+ \\
I_- \\
I_z
\end{matrix} \right] | \left[
\begin{matrix}
I_+ & I_- & I_z
\end{matrix} \right] ) = 
\left[ \begin{matrix}
\rbraket{I_+ | I_+} & 0 & 0 \\
0                   & \rbraket{I_- | I_-} & 0 \\
0                   & 0                   & \rbraket{I_z | I_z} 
\end{matrix} \right]
\label{eq:outerp2}
\end{equation}

\noindent and the inverse matrix $\rbraket{ \overrightarrow{ \mathscr{G} } |
\overleftarrow{\mathscr{G} }}^{-1}$ takes the diagonal form:

\begin{equation}
\left[ \begin{matrix}
\frac{1}{ \rbraket{I_+ | I_+} } & 0 & 0 \\
0 & \frac{ 1 }{ \rbraket{I_- | I_-} } & 0 \\
0 & 0 & \frac{ 1 }{ \rbraket{I_z | I_z} }
\end{matrix} \right].
\label{eq:outerp3}
\end{equation}

\noindent We note also that the entries of matrices
(\ref{eq:outerp2}) and (\ref{eq:outerp3}) are $c$-numbers and
therefore commute with any linear superoperator.

\subsection{Derivation of Mori's Equation}

The projector method separates the Liouville space into two subsets:

\begin{enumerate}
\item Deterministic and slow variables,
\item Rapid random fluctuations.
\end{enumerate}

These two orthogonal subspaces of the Liouville space are decomposed
using projectors as follows: $\pi$ is the projector onto the
deterministic variables while $1-\pi$ is the projector onto the random
part.

The usual treatment~\cite{bib:zwanzig_book,bib:kivelson,bib:kubo_toda2}
begins from the Dyson formula (Eq. ~\ref{eq:dyson1}),

\begin{equation}
 \mathrm{e}^{t\mathtensor{L}} = \mathrm{e}^{ t(1-\pi)
  \mathtensor{L}} + \int_0^t \mathrm{e}^{(t-s)
  \mathtensor{L}} \pi \mathtensor{L} \mathrm{e}^{s(1-\pi)
  \mathtensor{L}} \mathrm{d}s.
\label{eq:identityeL}
\end{equation}

We operate with this expression on the column vector $(1-\pi)
\mathtensor{L} \rket{ \overrightarrow{ \mathscr{G} } }$. For the left
hand side of Eq.(\ref{eq:identityeL}), $\mathrm{e}^{t\mathtensor{L}}
(1-\pi) \mathtensor{L} \rket{ \overrightarrow{\mathscr{G}} }$, we get

\begin{multline}
\left( \frac{\partial }{\partial
  t} \mathrm{e}^{t\mathtensor{L}} \right)
 \rket{ \overrightarrow{\mathscr{G}} } - 
\mathrm{e}^{t\mathtensor{L}}  \rbraket{ \mathtensor{L}
  \overrightarrow{\mathscr{G}} | \overleftarrow{\mathscr{G}}} \cdot 
 \rbraket{ \overrightarrow{\mathscr{G}} |
  \overleftarrow{\mathscr{G}}}^{-1}
\rket{ \overrightarrow{\mathscr{G}} }
\\
= \frac{\partial }{\partial t} \rket{ \overrightarrow{\mathscr{G}}
 (t)} - \rbraket{ \mathtensor{L}
  \overrightarrow{\mathscr{G}} | \overleftarrow{\mathscr{G}}} \cdot 
 \rbraket{ \overrightarrow{\mathscr{G}} |
  \overleftarrow{\mathscr{G}}}^{-1}
\rket{ \overrightarrow{\mathscr{G}}(t) },
\label{eq:eq17}
\end{multline}

\noindent where $\rket{ \overrightarrow{\mathscr{G}} (t) }$ is
shorthand for $\mathrm{e}^{t\mathtensor{L}} \rket{
  \overrightarrow{\mathscr{G}} }$. For the right hand side of
Eq.(\ref{eq:identityeL})

$$  \left[ \mathrm{e}^{ t(1-\pi) \mathtensor{L}} + \int_0^t
  \mathrm{e}^{(t-s) \mathtensor{L}} \pi \mathtensor{L}
  \mathrm{e}^{s(1-\pi) \mathtensor{L}} \mathrm{d}s \right] (1-\pi) \mathtensor{L}
 \rket{ \overrightarrow{\mathscr{G}} } $$

\noindent we have:

\begin{align}
 & \mathrm{e}^{ t(1-\pi)
 \mathtensor{L}} (1-\pi)
 \mathtensor{L} \rket{ \overrightarrow{\mathscr{G}} } + \nonumber \\
  & \int_0^t \mathrm{e}^{(t-s) \mathtensor{L}} \pi
 \mathtensor{L} \mathrm{e}^{s(1-\pi) \mathtensor{L}} 
 (1-\pi) \mathtensor{L} \rket{ \overrightarrow{\mathscr{G}}
 } \mathrm{d}s \nonumber \\
=& \rket{ \overrightarrow{\mathscr{F}} (t)} + \int_0^t \mathrm{e}^{(t-s)
 \mathtensor{L}} \pi \mathtensor{L} \rket{
 \overrightarrow{\mathscr{F}} (s)} \mathrm{d}s \nonumber \\
=& \rket{
 \overrightarrow{\mathscr{F}} (t)} + \int_0^t \mathrm{e}^{(t-s)
 \mathtensor{L}}
  \rbraket{  \mathtensor{L}
  \overrightarrow{\mathscr{F}}(s) | \overleftarrow{\mathscr{G}} }
  \cdot \rbraket{ \overrightarrow{\mathscr{G}} |
  \overleftarrow{\mathscr{G}} }^{-1} 
  \rket{ \overrightarrow{\mathscr{G}} } \mathrm{d}s \nonumber \\
 =& \rket{ \overrightarrow{\mathscr{F}} (t)} + \int_0^t \rbraket{
 \mathtensor{L}
 \overrightarrow{\mathscr{F}}(s) | \overleftarrow{\mathscr{G}} }
 \cdot \rbraket{ \overrightarrow{\mathscr{G}} |
 \overleftarrow{\mathscr{G}} }^{-1}  \rket{
 \overrightarrow{\mathscr{G}} (t-s)} \mathrm{d}s,
\label{eq:eq18}
\end{align}

\noindent where we have made the following abbreviation:

\begin{equation}
 \rket{ \overrightarrow{\mathscr{F}} (t) } = \mathrm{e}^{ t(1-\pi)
 \mathtensor{L}} (1-\pi)
\mathtensor{L} \rket{ \overrightarrow{\mathscr{G}} }.
\label{eq:randomforce}
\end{equation}

The last equality follows from the commutativity of the
operator $\mathrm{e}^{(t-s) \mathtensor{L}}$ with the matrices
$\rbraket{  \mathtensor{L} \overrightarrow{\mathscr{F}}(s) |
\overleftarrow{\mathscr{G}} }$ and $\rbraket{
\overrightarrow{\mathscr{G}} | \overleftarrow{\mathscr{G}} }^{-1}$
whose entries are $c$-numbers.

Combining the left and right hand sides together and making the
following definitions

\begin{align}
 \overleftrightarrow{\mathscr{K}} (s) =& \rbraket{  \mathtensor{L}
  \overrightarrow{\mathscr{F}}(s) | \overleftarrow{\mathscr{G}} }
  \cdot \rbraket{ \overrightarrow{\mathscr{G}} |
  \overleftarrow{\mathscr{G}} }^{-1}, \label{eq:memory} \\
i \overleftrightarrow{\Omega} =&
 \rbraket{ \mathtensor{L} \overrightarrow{\mathscr{G}} |
 \overleftarrow{\mathscr{G}}} \cdot \rbraket{
 \overrightarrow{\mathscr{G}} | \overleftarrow{\mathscr{G}}}^{-1},
\end{align}

\noindent we get the following set of coupled integro-differential
equations:

\begin{align}
 \frac{\partial }{\partial t} \rket{ \overrightarrow{\mathscr{G}} (t)}
 =& i \overleftrightarrow{\Omega}
 \rket{ \overrightarrow{\mathscr{G}} (t)} + \rket{
 \overrightarrow{\mathscr{F}} (t)} \nonumber \\
 & + \int_0^t \overleftrightarrow{\mathscr{K}}(s)
 \rket{ \overrightarrow{\mathscr{G}} (t-s)} \mathrm{d}s
\label{eq:mori}
\end{align}

\noindent which read, in component form:

\begin{align}
 \frac{\partial }{\partial t} \rket{\mathscr{G}_j (t)} =& \sum_k i\Omega_{jk}
 \rket{\mathscr{G}_k (t)} + \rket{ \mathscr{F}_j (t)} \nonumber \\
 & + \int_0^t \mathrm{d}s \sum_k \mathscr{K}_{jk} (s)
 \rket{\mathscr{G}_k (t-s)}.
\label{eq:mori_component}
\end{align}

These are the famous Mori-Zwanzig
equations
~\cite{bib:mori_fujisaka,bib:zwanzig,bib:kubo_toda2,bib:zwanzig_book}
describing the time-evolution of quantum operators in the presence
of lattice degrees of freedom.

\begin{itemize}
\item The first term on the right hand side is an instantaneous
oscillation term. It is a Markovian term, in the sense that it is evaluated
at the same time $t$ as the left hand side of the equation, regardless
of the past.

\item The second term, $\rket{ \mathscr{F}_j (t)}$, is a random driving
force because it is constructed using projectors $(1-\pi)$ onto the
``noise subspace'' of random variables, according to
Eq. (\ref{eq:randomforce}). This random force describes the effects of
the bath dynamics on the behavior of the spin system. 

\item The last term describes relaxation and is not necessarily
exponential; rather, it is a time-lagged memory term giving the
effects of past behavior on the present. Note that the quantity
$\rbraket{ \mathtensor{L} \overrightarrow{\mathscr{F}}(s) |
\overleftarrow{\mathscr{G}} }$ can be viewed as the covariance 
of $\mathtensor{L} \overrightarrow{\mathscr{F}}(s)$ and
the initial value $\overleftarrow{\mathscr{G}}$, assuming stationary
(in the wide sense) random processes.
\end{itemize}

\subsection{Weak Coupling Approximation}

Using definition of adjoint operator $(\mathtensor{L}^\dagger
\mathscr{P}, \mathscr{Q} ) = (\mathscr{P}, \mathtensor{L}
\mathscr{Q})$ and assuming an inner product is chosen such that
the Liouvillian is self-adjoint i.e. $\mathtensor{L}^\dagger
=\mathtensor{L}$, we can write the memory function
(Eq.~\ref{eq:memory}) as:

\begin{equation}
 \overleftrightarrow{\mathscr{K}}(s) = \rbraket{  \mathrm{e}^{
 s(1-\pi) \mathtensor{L}} (1-\pi) \mathtensor{L}
 \overrightarrow{\mathscr{G}} | \mathtensor{L}
 \overleftarrow{\mathscr{G}} } \cdot
 \rbraket{ \overrightarrow{\mathscr{G}} | \overleftarrow{\mathscr{G}}
 }^{-1}
\end{equation}

\noindent where 

$$C(s)=\rbraket{  \mathrm{e}^{
 s(1-\pi) \mathtensor{L}} (1-\pi) \mathtensor{L}
 \overrightarrow{\mathscr{G}} | \mathtensor{L}
 \overleftarrow{\mathscr{G}} }$$

\noindent is a time-correlation function.

Following Chorin~\cite{bib:chorin}, we consider two asymptotic
regimes with respect to the correlation time of the dynamics. This
correlation function contains the superoperator,

\begin{equation}
\pi \mathtensor{L} \mathrm{e}^{s(1-\pi) \mathtensor{L}} (1-\pi) \mathtensor{L}
\overrightarrow{\mathscr{G}}
\end{equation}

\noindent which remains unchanged if we insert the projector $(1-\pi)$:

\begin{equation}
\pi \mathtensor{L} (1-\pi) \mathrm{e}^{s(1-\pi) \mathtensor{L}} (1-\pi)
\mathtensor{L} \overrightarrow{\mathscr{G}}.
\end{equation}

\noindent This expression is also equal to

\begin{multline}
\pi \mathtensor{L} (1-\pi) \mathrm{e}^{s \mathtensor{L}} (1-\pi)
\mathtensor{L} \overrightarrow{\mathscr{G}} + \\
\pi \mathtensor{L} (1-\pi) \left[ \mathrm{e}^{s(1-\pi)} -
  \mathrm{e}^{s \mathtensor{L}} \right] \mathtensor{L} (1-\pi)
\mathtensor{L} \overrightarrow{\mathscr{G}}.
\label{eq:lhs1}
\end{multline}

\noindent Expanding the second term in a Taylor series:

\begin{align}
\mathrm{e}^{s(1-\pi)} - \mathrm{e}^{s \mathtensor{L}} =& \mathtensor{1} + s(1-\pi)
\mathtensor{L} + \dots - \mathtensor{1} - s\mathtensor{L} - \dots
\nonumber \\
 =& -s\pi \mathtensor{L} + O(s^2)
\end{align}

\noindent and since $(1-\pi) \pi = 0$, expression (\ref{eq:lhs1}) becomes:

\begin{equation}
\pi \mathtensor{L} (1-\pi) \mathrm{e}^{s \mathtensor{L}} (1-\pi)
\mathtensor{L} \overrightarrow{\mathscr{G}} + O(s^2).
\end{equation}

\noindent Integration from $0$ to $t$ yields

\begin{equation}
\int_0^t \pi \mathtensor{L} (1-\pi) \mathrm{e}^{s \mathtensor{L}} (1-\pi)
\mathtensor{L} \overrightarrow{\mathscr{G}} \mathrm{d}s + O(t^3).
\end{equation}

\noindent The term $O(t^3)$ can be neglected if the correlation time

\begin{multline}
\tau_c \cdot \mathtensor{1} = \rbraket{ (1-\pi) \mathtensor{L}
\overrightarrow{\mathscr{G}} | \mathtensor{L}
\overleftarrow{\mathscr{G}} }^{-1} \\
 \times \int_0^\infty \rbraket{  \mathrm{e}^{
s(1-\pi) \mathtensor{L}} (1-\pi) \mathtensor{L}
\overrightarrow{\mathscr{G}} | \mathtensor{L}
\overleftarrow{\mathscr{G}} } \mathrm{d}s
\end{multline}

\noindent corresponding to $C(t)$ is short. In this case, we may approximate
$\mathrm{e}^{s(1-\pi) \mathtensor{L}}$ by $\mathrm{e}^{ s
\mathtensor{L}}$ in the expression for the memory function. 

At the other extreme where $C(t)$ decays very slowly, the exponential
$\mathrm{e}^{s(1-\pi) \mathtensor{L}}$ does not vary appreciably and the
approximation 

\begin{equation}
\mathrm{e}^{s(1-\pi) \mathtensor{L}} = \mathrm{e}^{s \mathtensor{L}}
\end{equation}

\noindent holds (expand the memory function in powers of $s$ and retain
the zeroth order term). 

In both cases, this approximation reflects the fact
that the random fluctuations have negligible effect on the
time-evolution. This is called the weak-coupling
approximation~\cite{bib:chorin}.

In this approximation, the memory function is given by:

\begin{equation}
 \overleftrightarrow{\mathscr{K}}(s) = \rbraket{ (1-\pi) \mathrm{e}^{
 s \mathtensor{L}} (1-\pi) \mathtensor{L}
 \overrightarrow{\mathscr{G}} | \mathtensor{L}
 \overleftarrow{\mathscr{G}} } \cdot
 \rbraket{ \overrightarrow{\mathscr{G}} | \overleftarrow{\mathscr{G}}
 }^{-1}.
\end{equation}

\subsection{Short Correlation Times \label{sec:shortcorrel} }

\noindent The integral term,

\begin{equation}
 \int_0^t \mathrm{d}s \sum_k \mathscr{K}_{jk} (s) \rket{\mathscr{G}_k (t-s)}
\end{equation}

\noindent can be approximated by

\begin{equation}
 \rket{\mathscr{G}_k (t)} \int_0^\infty \mathrm{d}s \sum_k \mathscr{K}_{jk} (s) 
\end{equation}

\noindent if the correlation time $\tau_c$ is much shorter than the
observation time $t$. Such approximations are accurate, for example,
when the random fluctuations in the lattice are very
rapid ($\tau_c \sim 10^{-12} s$) compared to the observation time $t
\sim 10^{-3} s$.

\subsection{Strong Zeeman Fields}

In this section, we work out the strong field approximation needed to
derive the famous example of the NMR Bloch equations. To keep the
presentation simple, we make use of the inner product of
Eq.~(\ref{eq:ip1}). While this approach foregoes the concept of spin
temperature and equilibrium as is normally done in the standard 
treatments of NMR relaxation~\cite{bib:abragam}, it does simplify the
presentation and provides some relaxation rates. We
write~\cite{bib:ernstbook,bib:budker_book}

\begin{equation}
 \mathbf{F} = \sum_i \mathbf{I}_i 
\end{equation}

\noindent for the total nuclear spin angular momentum operator.
For a Hamiltonian $H$ which consists of a lattice part $H_L$, and a
spin part subdivided into Zeeman $H_0=\omega_0 F_z$ and small
perturbation $H_1$ ($\| H_1 \| \ll \| H_0\|$):

\begin{equation}
H = H_L + H_0 + H_1,
\end{equation}

\noindent the oscillation term for the transverse component
$F_{\pm} = F_x \pm i F_y$ of
$\mathscr{G}$ is

\begin{equation}
 \frac{1}{i} \frac{ \rBraket{ F_{\pm} |
  \mathtensor{L} | F_{\pm} } }{
  \rbraket{ F_{\pm} | F_{\pm} } }  =  \frac{
  \rBraket{ F_{\pm} |  [H_0,
  F_{\pm} ] } }{ \rbraket{ F_{\pm} |
  F_{\pm} } } + \frac{ \rBraket{ F_{\pm} |
  [H_1, F_{\pm} ] } }{
  \rbraket{ F_{\pm} | F_{\pm} } }.
\end{equation}

Using the commutator $[F_z, F_{\pm}]=\pm F_{\pm}$ and neglecting the
second term since $H_1$ is assumed small compared to
$H_0$ we are left with

\begin{equation}
\Omega_{\pm\pm} = \frac{ \rBraket{ F_{\pm} |  [\omega_0
 F_z, F_{\pm} ]}}{\rbraket{ F_{\pm} |
 F_{\pm} }} = \pm \omega_0.
\end{equation}

\noindent For the longitudinal component we have:

\begin{equation}
\Omega_{zz} = \frac{1}{i} \frac{\rBraket{ F_z |
 \mathtensor{L} | F_z }}{\rbraket{ F_z | F_z }} \approx \frac{
 \rBraket{ F_z | [H_0, F_z ]}}{\rbraket{ F_z
 | F_z }} =0.
\end{equation}

Thus, the $\overleftrightarrow{\Omega}$ matrix for the set of
operators $\{ F_+, F_-, F_z\}$ is

$$\overleftrightarrow{\Omega} =
\left[ \begin{matrix}
\Omega_{++} & \Omega_{+-} & \Omega_{+z} \\
\Omega_{-+} & \Omega_{--} & \Omega{-z} \\
\Omega_{z+} & \Omega_{z-} & \Omega{zz}
\end{matrix} \right]=
\left[ \begin{matrix}
+\omega_0 & 0 & 0 \\
0 & -\omega_0 & 0 \\
0 & 0 & 0
\end{matrix} \right].$$

Omitting the random driving force and using these approximations for
$\Omega$, the Mori master equations for $F_\pm$ and
$F_z$ read

\begin{align}
 \frac{\partial }{\partial t} \rket{ F_\pm(t)} =& \pm i\omega_0
 \rket{ F_\pm(t)} \nonumber \\
 & + \int_0^\infty \mathrm{d}s \sum_{j=+,-,z}  K_{\pm,j}(s)
 \rket{ F_j(t-s)} \nonumber \\
 \frac{\partial }{\partial t} \rket{ F_z (t)} =& \int_0^\infty \mathrm{d}s
 \sum_{j=+,-,z} K_{z,j}(s) \rket{ F_j (t-s)}
\end{align}

\noindent where $K_{ij}(s)$ is the $3 \times 3$ matrix whose elements
are

\begin{equation}
 K_{ij}(s) = \frac{ \rBraket{ F_i | \mathtensor{L} \mathrm{e}^{ s
 \mathtensor{L}} (1-\pi) \mathtensor{L} | F_j } }{ \rbraket{F_j | F_j} }.
\end{equation}

These equations involve integral operators with past information.
Instead of scalar relaxation rates $1/T_1$ and $1/T_2$ they contain
convolution integrals over the memory functions
$K_{ij}(s)$. Consequently, spin relaxation is not necessarily
exponential.

\subsection{Bloch Equations}

To recover the classical Bloch equations, we make the following
assumptions:

\begin{enumerate}
\item The memory function is a diagonal matrix. We will return to this
  point shortly.
\item The correlation time is short (see
Section~\ref{sec:shortcorrel})
\item The set of operators is still $\{ F_+, F_-, F_z \}$, however, in
 the equations of motion, we substitute the deviations from thermal
 equilibrium
$$ \mathbf{F}(t) \rightarrow \mathbf{F}(t) - \langle \mathbf{F}
 \rangle_{eq}. $$
\end{enumerate}

\noindent We then obtain the classical Bloch equations

\begin{align}
 \frac{\partial }{\partial t} \rket{ F_\pm(t) } =& \pm i\omega_0
 \rket{ F_\pm(t) }  - \frac{ \rket{ F_\pm(t)} }{ T_2 },  \nonumber \\
\frac{\partial }{\partial t} \rket{ F_z (t) } =& - \frac{ \rket{ F_z
(t) } - \langle F_z \rangle_{eq} }{T_1},
\label{eq:classbloch}
\end{align}

\noindent where the relaxation rates are

\begin{align}
\frac{1}{T_2} =& - \int_0^\infty \frac{ \rBraket{ \mathtensor{L} F_\pm | 
 \mathrm{e}^{ s \mathtensor{L}} | (1-\pi) \mathtensor{L}  F_\pm } }{
 \rbraket{F_\pm | F_\pm } } \mathrm{d}s, \nonumber \\
\frac{1}{T_1} =&  -\int_0^\infty \frac{ \rBraket{ \mathtensor{L} F_z | 
 \mathrm{e}^{ s \mathtensor{L}} | (1-\pi) \mathtensor{L} F_z } }{
 \rbraket{ F_z | F_z } } \mathrm{d}s.
\label{eq:relaxrates}
\end{align}

The memory function is diagonal if there are no couplings. The
projection superoperator

\begin{equation}
\pi \rket{ \mathtensor{L} \overrightarrow{ F } } = \rbraket{
  \mathtensor{L} \overrightarrow{ F } | \overleftarrow{ F }} \cdot \rbraket{
  \overrightarrow{ F } | \overleftarrow{ F }}^{-1} \rket{
  \overrightarrow{ F } },
\end{equation}

\noindent takes the following form

\begin{equation}
 \left[ \begin{matrix}
 \frac{ \rbraket{ \mathtensor{L} F_+ | \mathtensor{L} F_+  }}{
 \rbraket{F_+ | F_+} } & \frac{ \rbraket{ \mathtensor{L} F_+ |
 \mathtensor{L} F_-  } }{ \rbraket{F_- | F_-} } & \frac{ \rbraket{
 \mathtensor{L} F_+ | \mathtensor{L} F_z  }}{ \rbraket{F_z | F_z} } \\
 \frac{ \rbraket{ \mathtensor{L} F_- | \mathtensor{L} F_+  } }{
 \rbraket{F_+ | F_+} } & \frac{ \rbraket{ \mathtensor{L} F_- |
 \mathtensor{L} F_-  } }{ \rbraket{F_- | F_-} } & \frac{ \rbraket{
 \mathtensor{L} F_- | \mathtensor{L} F_z  } }{ \rbraket{F_z | F_z} } \\
 \frac{ \rbraket{ \mathtensor{L} F_z | \mathtensor{L} F_+  } }{
 \rbraket{F_+ | F_+} } & \frac{ \rbraket{ \mathtensor{L} F_z |
 \mathtensor{L} F_-  } }{ \rbraket{F_- | F_-} } & \frac{ \rbraket{
 \mathtensor{L} F_z | \mathtensor{L} F_z  } }{ \rbraket{F_z | F_z} }
\end{matrix} \right]
 \rket{  \left[
\begin{matrix}
F_+ \\
F_- \\
F_z
\end{matrix} \right] }.
\end{equation}

In the case of no couplings, $H = H_L + H_Z$, and off-diagonal elements
such as 

$$\rbraket{ \mathtensor{L} F_+ | \mathtensor{L} F_-  } =
-\rbraket{ F_+ | F_- }$$

\noindent vanish due to the orthogonality of the $F_i$'s.
The matrix then simplifies to

\begin{equation}
 \left[ \begin{matrix}
 \frac{ \rbraket{ \mathtensor{L} F_+ | \mathtensor{L} F_+  }}{
 \rbraket{F_+ | F_+} } & 0 & 0 \\
 0 & \frac{ \rbraket{ \mathtensor{L} F_- | \mathtensor{L} F_-  } }{
 \rbraket{F_- | F_-} } & 0 \\
 0 & 0 & \frac{ \rbraket{ \mathtensor{L} F_z | \mathtensor{L} F_z  }
 }{ \rbraket{F_z | F_z} }
\end{matrix} \right]
 \rket{  \left[
\begin{matrix}
F_+ \\
F_- \\
F_z
\end{matrix} \right] }.
\label{eq:diagmatrix}
\end{equation}

When couplings are weak ($\| H_D \| \ll \| H_Z \|$), these simplified
equations of motion [uncoupled Bloch equations,
Eq.(\ref{eq:classbloch})] yield a reasonable approximation.

\section{Time-Dependent Hamiltonians}

We now show how the previous treatment can be extended to accomodate
time-dependent Hamiltonians. 

\subsection{Time-Evolution Operator}

The starting point is the Heisenberg equation of motion for an
observable $A(t)$:

\begin{equation}
\frac{d}{dt} A(t) = \mathtensor{L}(t) A(t)
\label{eq:eq11}
\end{equation}

\noindent where $\mathtensor{L}(t)$ is a time-dependent
Liouvillian defined by

\begin{equation}
\mathtensor{L}(t) A(t) = i [ H(t), A(t) ].
\end{equation}

\noindent If we write the solution in the form

\begin{equation}
 A(t) = U(0,t) A(0)
\label{eq:eq2}
\end{equation}

\noindent then Eq.~(\ref{eq:eq11}) implies:

\begin{equation}
\frac{d}{dt} A(t) = \frac{ dU }{dt} A(0) =
\mathtensor{L}(t) U(0,t) A(0)
\end{equation}

\noindent and since this relation holds for arbitrary initial
conditions $A(0)$, we have the general formula which defines the
time-evolution operator:

\begin{equation}
\frac{ dU(0,t) }{ dt} = \mathtensor{L}(t) U(0,t).
\label{eq:eq4}
\end{equation}

The solution of this equation is given by:

\begin{widetext}
\begin{equation}
U(0,t) = \sum_{n=0}^\infty \int_0^t d\tau_1 \int_0^{\tau_1} d\tau_2
\dots \int_0^{\tau_{n-1}} d\tau_{n} \mathtensor{L}(\tau_1)
\mathtensor{L}(\tau_2) \dots \mathtensor{L}(\tau_n)
\label{eq:eq5}
\end{equation}
\end{widetext}

\noindent where the Liouvillians are {\it left time
ordered}. Consequently, Eq.~(\ref{eq:eq11}) is a {\it
causal} relation. That Eq.~(\ref{eq:eq5}) is the solution of
Eq.~(\ref{eq:eq4}) can be seen by differentiating Eq.~(\ref{eq:eq5})
with respect to $t$:

\begin{widetext}
\begin{equation}
\frac{dU}{dt} = \mathtensor{L}(t) \sum_{n=0}^\infty \int_0^t
d\tau_1 \int_0^{\tau_1} d\tau_2
\dots \int_0^{\tau_{n-1}} d\tau_{n} \mathtensor{L}(\tau_1)
\mathtensor{L}(\tau_2) \dots \mathtensor{L}(\tau_n).
\label{eq:eq6}
\end{equation}
\end{widetext}

\noindent A shorthand notation for $U(0,t)$ uses the {\it left}
Dyson time-ordering operator $\overleftarrow{T}$:

\begin{equation}
U(0,t) = \overleftarrow{T} \exp \left( \int_0^t \mathtensor{L}(\tau)
d\tau \right)
\end{equation}

\noindent For general time evolution, we have:

\begin{widetext}
\begin{equation}
U(s,t) = \sum_{n=0}^\infty \int_s^t d\tau_1 \int_s^{\tau_1} d\tau_2
\dots \int_s^{\tau_{n-1}} d\tau_{n} \mathtensor{L}(\tau_1)
\mathtensor{L}(\tau_2) \dots \mathtensor{L}(\tau_n)
\end{equation}
\end{widetext}

\noindent which corresponds to

\begin{equation}
U(s,t) = \overleftarrow{T} \exp \left( \int_s^t \mathtensor{L}(\tau)
d\tau \right).
\end{equation}

\subsection{Dyson Formula for Time-Dependent Liouvillian}

Now we prove the following identity, which holds, for time-dependent
Liouvillian,

\begin{widetext}
\begin{equation}
 \overrightarrow{T} \exp \left( \int_0^t d\tau \mathtensor{L}(\tau) \right)
 = \overrightarrow{T} \exp \left( \int_0^t d\tau (1-\pi) \mathtensor{L}(\tau)
 \right) + \int_0^t  \overrightarrow{T} \exp \left( \int_0^s d\tau
 \mathtensor{L}(\tau) \right) \pi \mathtensor{L}(s)  \overrightarrow{T}
 \exp \left( \int_s^t d\tau (1-\pi) \mathtensor{L}(\tau) \right) \mathrm{d}s,
\label{eq:dyson2}
\end{equation}
\end{widetext}

\noindent where the {\it right time ordering} operator is defined by
the time arguments increasing as we go from the left to the right

\begin{widetext}
\begin{equation}
 \overrightarrow{T} \exp \left( \int_s^t d\tau \mathtensor{L}(\tau)
 \right) = \sum_{n=0}^\infty \int_s^t d\tau_1 \int_s^{\tau_1} d\tau_2
\dots \int_s^{\tau_{n-1}} d\tau_{n}  \mathtensor{L}(\tau_n)
 \mathtensor{L}(\tau_{n-1}) \dots \mathtensor{L}(\tau_1).
\end{equation}
\end{widetext}

\noindent We first rewrite Eq.~(\ref{eq:dyson2}) as
\begin{widetext}
\begin{align}
 - \overrightarrow{T} \exp \left( \int_0^t d\tau (1-\pi) \mathtensor{L}(\tau)
 \right) =& - \overrightarrow{T} \exp \left( \int_0^t d\tau
 \mathtensor{L}(\tau) \right)
 + \nonumber \\
 & \quad \int_0^t  \overrightarrow{T} \exp \left( \int_0^s d\tau
 \mathtensor{L}(\tau) \right) \pi \mathtensor{L}(s)  \overrightarrow{T}
 \exp \left( \int_s^t d\tau (1-\pi) \mathtensor{L}(\tau) \right) \mathrm{d}s,
\end{align}
\end{widetext}

\noindent and show that each side of the equality obeys the same
differential equation for the same initial conditions. For the left
hand side, we have:

\begin{equation}
\frac{d}{dt} LHS = LHS \cdot (1-\pi) \mathtensor{L}(t).
\end{equation}

\noindent While for the right hand side:

\begin{widetext}
\begin{align}
\frac{d}{dt} RHS =& - \overrightarrow{T} \exp
\left( \int_0^t d\tau \mathtensor{L}(\tau) \right) \mathtensor{L}(t) + 
 \overrightarrow{T} \exp \left( \int_0^t d\tau \mathtensor{L}(\tau)
 \right) \pi \mathtensor{L}(t) \nonumber \\
 & + \int_0^t \overrightarrow{T} \exp \left( \int_0^s d\tau
 \mathtensor{L}(\tau) \right) \pi \mathtensor{L}(s)  \overrightarrow{T}
 \exp \left( \int_s^t d\tau (1-\pi) \mathtensor{L}(\tau) \right) \mathrm{d}s
 (1-\pi) \mathtensor{L}(t) \nonumber \\
 =& RHS \cdot (1-\pi) \mathtensor{L}(t)
\end{align}
\end{widetext}

These identical differential equations, combined with the
initial condition at $t=0$, complete the proof of the Dyson formula for
time-dependent Liouvillian. This identity 
is due to Holian and Evans ~\cite{bib:holian_evans,bib:evans_morriss}.

\subsection{Derivation of the Mori Formulae Using Time-Dependent Dyson
Formula}

Using Eq.~(\ref{eq:dyson2}) we may derive the Mori equations in a
manner analogous to the time-independent case. The analogue of
Eq.~(\ref{eq:eq17}) is the following: let the left
hand side of Eq.~(\ref{eq:dyson2}) act on $(1-\pi) \mathtensor{L}(t)
\rket{ \overrightarrow{ \mathscr{G} } }$,

\begin{equation}
 \frac{\partial }{\partial t} \rket{ \overrightarrow{
 \mathscr{G}}^\dagger (t)} - \rbraket{ \mathtensor{L}(t)
  \overrightarrow{\mathscr{G}} | \overleftarrow{\mathscr{G}}} \cdot 
 \rbraket{ \overrightarrow{\mathscr{G}} |
  \overleftarrow{\mathscr{G}}}^{-1}
\rket{ \overrightarrow{\mathscr{G}}^\dagger (t) }
\end{equation}

\noindent where 

\begin{equation}
\mathscr{G}^\dagger (t) \equiv \overrightarrow{T} \exp \left( \int_0^t
d\tau \mathtensor{L}(\tau) \right) \mathscr{G}
\end{equation}

\noindent is the {\it anti-causal} propagation of $\mathscr{G}$. Next, we
let the right hand side of Eq.~(\ref{eq:dyson2}) act on $(1-\pi)
\mathtensor{L}(t) \rket{ \overrightarrow{ \mathscr{G} } }$, and obtain
an expression analogous to Eq.~(\ref{eq:eq18}):

\begin{equation}
 \rket{ \overrightarrow{\mathscr{F}}^\dagger (0,t)} + \int_0^t
 \rbraket{ \mathtensor{L}(s) \overrightarrow{\mathscr{F}}^\dagger
 (s,t) | \overleftarrow{\mathscr{G}} }
 \cdot \rbraket{ \overrightarrow{\mathscr{G}} |
 \overleftarrow{\mathscr{G}} }^{-1}  \rket{
 \overrightarrow{\mathscr{G}}^\dagger (s)} \mathrm{d}s
\end{equation}

\noindent where

\begin{equation}
 \rket{ \overrightarrow{\mathscr{F}}^\dagger (s,t) } \equiv
 \overrightarrow{T} \exp \left(
 \int_s^t d\tau (1-\pi) \mathtensor{L}(\tau) \right) (1-\pi)
 \mathtensor{L}(t) \rket{ \overrightarrow{\mathscr{G}} }.
\label{eq:a1}
\end{equation}

\noindent Combining these expressions and making the following
definitions

\begin{align}
 \overleftrightarrow{\mathscr{K}}^\dagger (s,t) \equiv & \rbraket{
 \mathtensor{L}(s)  \overrightarrow{\mathscr{F}}^\dagger(s,t) |
 \overleftarrow{\mathscr{G}} }
  \cdot \rbraket{ \overrightarrow{\mathscr{G}} |
  \overleftarrow{\mathscr{G}} }^{-1}, \label{eq:a2} \\
i \overleftrightarrow{\Omega}(t) \equiv &
 \rbraket{ \mathtensor{L}(t) \overrightarrow{\mathscr{G}} |
 \overleftarrow{\mathscr{G}}} \cdot \rbraket{
 \overrightarrow{\mathscr{G}} | \overleftarrow{\mathscr{G}}}^{-1},
\end{align}

\noindent we get the following set of coupled integro-differential
equations

\begin{align}
 \frac{\partial }{\partial t} \rket{ \overrightarrow{\mathscr{G}}^\dagger (t)}
 =& i \overleftrightarrow{\Omega}(t)
 \rket{ \overrightarrow{\mathscr{G}}^\dagger (t)} + \rket{
 \overrightarrow{\mathscr{F}}^\dagger (0,t)} \nonumber \\
 & + \int_0^t \overleftrightarrow{\mathscr{K}}^\dagger(s,t)
 \rket{ \overrightarrow{\mathscr{G}}^\dagger (s)} \mathrm{d}s,
\end{align}

\noindent which read in component form

\begin{align}
 \frac{\partial }{\partial t} \rket{\mathscr{G}_j^\dagger (t)} =&
 \sum_k i\Omega_{jk}(t) \rket{\mathscr{G}_k^\dagger (t)} + \rket{
 \mathscr{F}_j^\dagger (0,t)} \nonumber \\
 & + \int_0^t \mathrm{d}s \sum_k \mathscr{K}_{jk}^\dagger (s,t)
 \rket{\mathscr{G}_k^\dagger (s)}.
\label{eq:a3}
\end{align}

These are the Mori equations for time-dependent Liouvillian. It is
noteworthy that the memory kernel $\mathscr{K}_{jk}^\dagger (s,t)$
no longer depends on the time difference $t-s$, but also depends on
the origin of time ($s$). This is a consequence of the
non-commutativity of the Hamiltonian at different points in time.

Another point of interest is that the dynamics are given for the
anti-causal operators. An adjoint equation of motion can
also be derived which describes causal evolution, but requires adjoint
definitions of the propagator and the expression upon which the Dyson
equation is applied to. The weak coupling approximation of the
previous sections remains applicable, if it is needed.

We also note that the Liouvillian which appears under the integral
sign of Eq.(\ref{eq:a3}) cannot be replaced by an average
Liouvillian (see Eqs.~\ref {eq:a1} and \ref{eq:a2}). Such a
substitution is not generally justified. However, it is possible to
replace the exponential function of Eq.(\ref{eq:a1}) by a Magnus
expansion.

\subsection{Change of Reference Frames}

For convenience, we may change frames of reference at will. This is
done using time-ordered transformations, as is normally done when
solving the Liouville von Neumann equation. We summarize these
well-known transformations in the notation of this article.

\subsubsection{Interaction Representation}

If the time-dependent Liouvillian decomposes into the form
$\mathtensor{L}(t) = \mathtensor{L}_0(t) + \mathtensor{L}_V(t)$, the
Heisenberg equation of motion is

\begin{equation}
 \frac{dA(t)}{dt} = \bigl[ \mathtensor{L}_0(t) +
\mathtensor{L}_V(t) \bigr] A(t).
\end{equation}

\noindent We define the transformed operators:

\begin{align}
 \tilde{A} =& \overrightarrow{T} \exp \left( - \int_0^t d\tau
 \mathtensor{L}_0(\tau) \right) A, \\
 \tilde{\mathtensor{L}} =& \overrightarrow{T} \exp \left( - \int_0^t d\tau
 \mathtensor{L}_0(\tau) \right) \mathtensor{L},
\end{align}

\noindent where $\overrightarrow{T}$ indicates a time-ordering from
left to right, with the most recent Liouvillian to the right. These
transformations are {\it anti-causal}. Then,

\begin{widetext}
\begin{align}
 \frac{ d \tilde{A} }{dt} =& \frac{d}{dt}  \overrightarrow{T} \exp
 \left( - \int_0^t d\tau \mathtensor{L}_0(\tau) \right) A \nonumber \\
 =& - \overrightarrow{T} \exp\left( - \int_0^t d\tau
 \mathtensor{L}_0(\tau) \right) \mathtensor{L}_0(t) A +
 \overrightarrow{T} \exp\left( -
 \int_0^t d\tau \mathtensor{L}_0(\tau) \right)
 \bigl[ \mathtensor{L}_0(t) A +
 \mathtensor{L}_V(t) A(t) \bigr] \nonumber \\
 =& \tilde{\mathtensor{L}}_V(t) \tilde{A}.
\label{eq:interrp}
\end{align}
\end{widetext}

Thus, in a rotating frame of reference, a Liouville equation

\begin{equation}
\frac{d \tilde{A}}{dt} = \tilde{\mathtensor{L}}_V \tilde{A}
\end{equation}

\noindent holds, where $\tilde{\mathtensor{L}}_V$ is the perturbation
part of the Liouvillian expressed in the rotating frame.

\subsubsection{Toggling Frame Transformation}

In Equation (\ref{eq:interrp}), we may further decompose the
interaction representation Liouvillian

\begin{equation}
\tilde{\mathtensor{L}}_V = \tilde{\mathtensor{L}}_1 +
\tilde{\mathtensor{L}}_2,
\end{equation}

\noindent where $\tilde{\mathtensor{L}}_1$ is the
part corresponding to the RF field and
$\tilde{\mathtensor{L}}_2$ is the remaining ``internal''
part of the Liouvillian (e.g. spin-spin
interactions).

Consider a transformation to the so-called ``toggling frame'',

\begin{equation}
 \hat{\tilde{A}} = \overrightarrow{T} \exp \left( - \int_0^t
 \tilde{\mathtensor{L}}_1(t') dt' \right) \tilde{A},
\end{equation}

\noindent where $\overrightarrow{T}$ is the Dyson time-ordering
operator~\cite{bib:ernstbook}, which places the most recent events to
the right. Then,

\begin{align}
\frac{ d \hat{\tilde{A}}(t) }{dt} = \hat{\tilde{\mathtensor{L}}}_2
\hat{\tilde{A}}(t).
\end{align}

The solution to this equation is,

\begin{equation}
\hat{\tilde{A}}(t) = \overrightarrow{T} \exp \left( \int_0^t
 \hat{\tilde{\mathtensor{L}}}_2 (t') dt' \right)
 \hat{\tilde{A}}(0),
\end{equation}

\noindent where

\begin{equation}
\hat{\tilde{\mathtensor{L}}}_2 = \overrightarrow{T} \exp \left(- \int_0^t 
\tilde{\mathtensor{L}}_1 (t') dt' \right) \tilde{\mathtensor{L}}_2,
\end{equation}

\noindent and

\begin{equation}
\tilde{\mathtensor{L}}_2 = \overrightarrow{T} \exp \left( - \int_0^t d\tau
\mathtensor{L}_0(\tau) \right) \mathtensor{L}_2.
\end{equation}

\section{Conclusion}

We have extended the Mori theory to the case of time-dependent
Hamiltonians. To the author's knowledge, the previously published
treatments of Mori relaxation theory in the
NMR literature did not include the case of time-dependent
Liouvillian. This makes it possible to treat general cases of cw or
pulsed irradiations in NMR and/or coherent optics, for arbitrary
timescales of the molecular motions relative to the period of
irradiation. Also, we have outlined a way of introducing
coherent averaging effects into the Mori equations.

\bibliographystyle{unsrt}
\bibliography{dbase9}

\end{document}